\documentclass[]{spie}  

\usepackage[]{graphicx}

\title{Composition of the Chandra ACIS contaminant} 

\author{Herman L. Marshall\supit{a},
Allyn Tennant\supit{b}, \\
Catherine E. Grant\supit{a},
Adam P. Hitchcock\supit{c},
Steve O'Dell\supit{b},
and Paul P. Plucinsky\supit{d}
\skiplinehalf
\supit{a}Center for Space Research, Massachusetts Institute of Technology,
Cambridge, MA, USA 02139 \\
\supit{b}NASA Marshall Space Flight Center,
Huntsville, AL, USA 35812 \\
\supit{c}Chemistry, McMaster University, Hamilton, ON, Canada L8S-9140 \\
\supit{d}Harvard-Smithsonian Center for Astrophysics,
60 Garden St, Cambridge, MA, USA 02138 \\
}

\authorinfo{Further author information: (Send correspondence to
H.L.M.)\\H.L.M.: E-mail: hermanm@space.mit.edu, Telephone: 1 617 253 8573}

\begin{document} 
  \maketitle 

\newcommand\arcdeg{\mbox{$^\circ$}}%
\newcommand\arcmin{\mbox{$^\prime$}}%
\newcommand\arcsec{\mbox{$^{\prime\prime}$}}%
\let\jnlstyle=\rmfamily
\def\refj#1{{\jnlstyle#1}}%
\newcommand\pasp{\refj{PASP}}%

\begin{abstract}
The Advanced CCD Imaging Spectrometer (ACIS) on the Chandra X-ray
Observatory is suffering a gradual loss of low energy sensitivity due to
a buildup of a contaminant.  High resolution spectra of bright
astrophysical sources using the Chandra Low Energy Transmission Grating
Spectrometer (LETGS) have been analyzed in order to determine the nature
of the contaminant by measuring the absorption edges.  The dominant
element in the contaminant is carbon.  Edges due to oxygen and fluorine
are also detectable.
Excluding H, we find that C, O,
and F comprise $>$80\%, 7\%, and 7\% of the contaminant by number,
respectively.  Nitrogen is less than 3\% of the contaminant.  We will
assess various candidates for the contaminating material and
investigate the growth of the layer with time.
For example, the detailed structure of the C-K absorption edge
provides information about the bonding structure of the compound,
eliminating aromatic hydrocarbons as the contaminating material.

\end{abstract}



\section{Introduction}
\label{sect:intro}  

Plucinsky {\it et al.} (2003)\cite{plucinsky03}
showed that the detection efficiency of the {\em Chandra}\cite{weisskopf2002}
Advanced CCD Imaging Spectrometer (ACIS)\cite{garmire2003}
declined monotonically in the first
three years.  The decline was particularly apparent in the strengths
of the Mn and Fe L lines relative to the
Mn K line from the external calibration source; this ratio
dropped by 35\%.  The decline was attributed to the buildup of a contaminating
layer on the ACIS optical blocking filter, which is colder than most
components of the spacecraft interior.  Plucinsky
{\it et al.} examined data
from spectroscopy using the {\em Chandra} Low Energy Transmission Grating
Spectrometer (LETGS) used with the ACIS detector to show that the contaminant
had a very strong C-K edge and a possible weak O-K edge as well.  At the time,
there was no detection of the N-K edge and there was only a hint of an edge
at F-K.

In this paper, we construct a detailed empirical model of the absorption
due to the contaminant using observations of bright continuum-dominated
sources taken with the {\em Chandra} LETGS and ACIS.  We also describe
new features of this absorption model and the detection of a F-K edge.
A time dependent absorption model is constructed that can be used to
correct {\em Chandra} grating spectra for absorption.  This model is
compared to the results from the external calibration source.  We find that there
is still a component of the absorption that is unaccounted for in the new
absorption model that can cause systematic errors in the effective area
of up to 15\%.  These results rely on calibration analyses outlined by
Marshall {\it et al.} (2003)\cite{marshall03}.

\section{Early Results}

\begin{figure}
   \begin{center}
   \begin{tabular}{c}
   \includegraphics[angle=90,height=10cm]{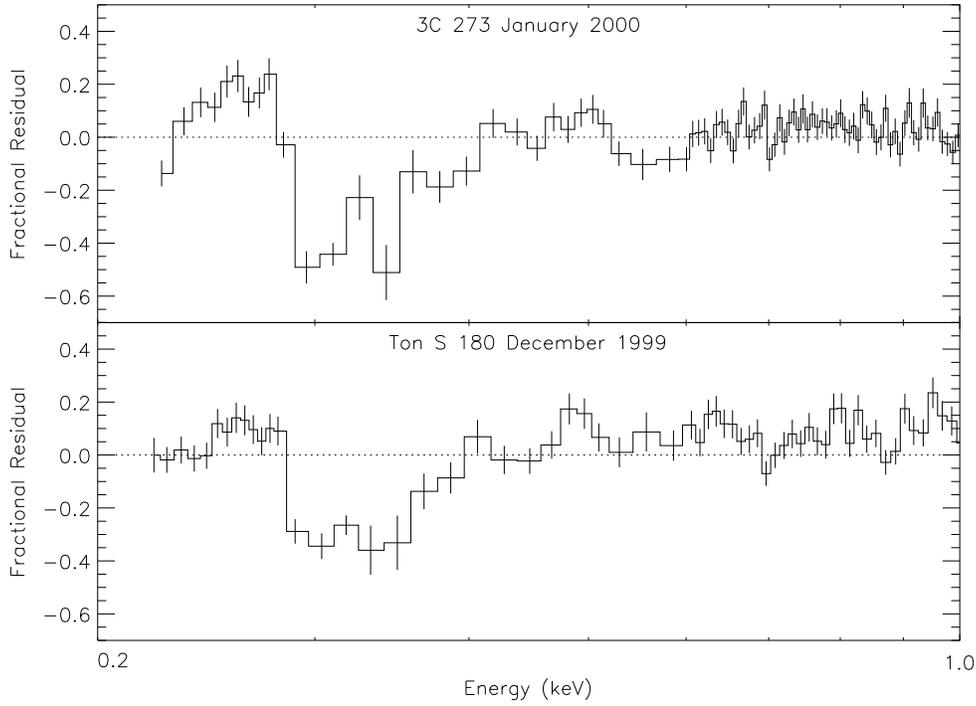}
   \end{tabular}
   \end{center}
   \caption
   { \label{fig:2src}
Comparison of two LETG/ACIS-S observations of bright continuum
sources.Ê The residuals
give the fractional differences between the simple power law
spectral model that was folded through the instrument response and
the observed count spectra.Ê Note the apparent edge at about 0.29 keV,
which is near the K edge of carbon.  The data were adaptively binned
to obtain a signal to noise ratio of at lest 20 in each bin,
limited to a bin width $<$ 5\%$E$ at energy $E$.}
   \end{figure}

The problem was first noticed in a Guest Observer observation (see
Fig.\ref{fig:2src}) of Ton S 180 taken in December 1999 (observation
ID 811).
This feature was subsequently detected in the calibration
observation of 3C 273 (taken in January 2000;
observation ID 1198). The residuals of the Ton S 180
and 3C 273 are compared in Fig.\ref{fig:2src}. The similarity is
quite striking and indicated that the difference was intrinsic to
the instrument.
The LETG/HRC-S data from 3C 273
taken immediately prior to the LETG/ACIS-S observation do not show
a comparable residual, confirming that the feature is not
intrinsic to the source.

\begin{figure}
   \begin{center}
   \begin{tabular}{c}
   \includegraphics[angle=90,height=10cm]{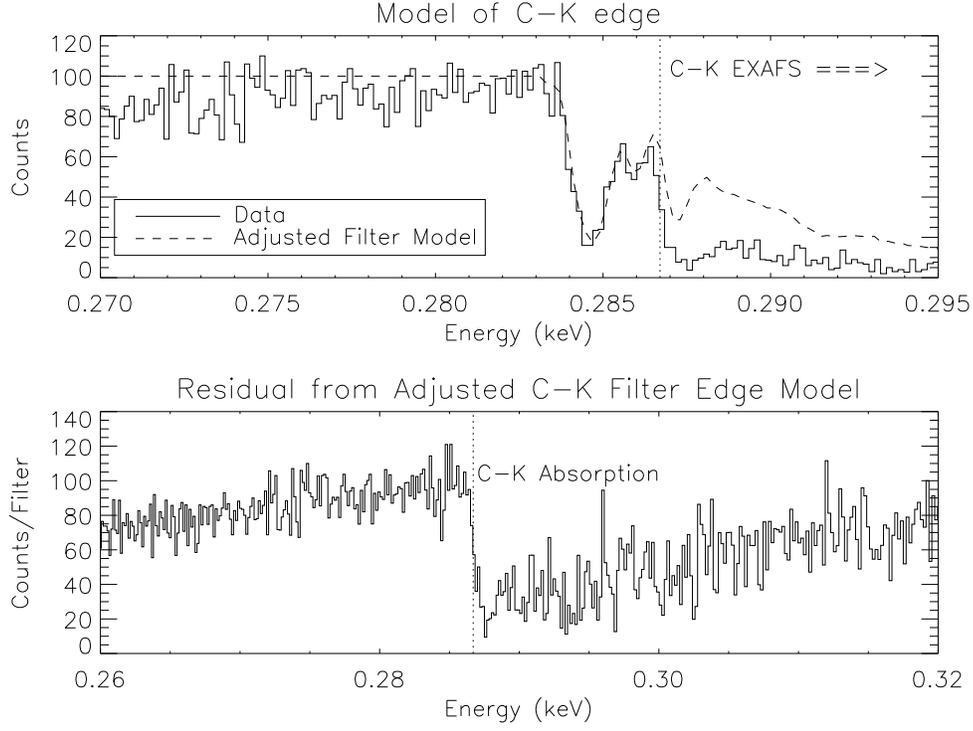}
   \end{tabular}
   \end{center}
   \caption
{ \label{fig:xtej1118}
Modelling of the C-K edge region using the
data from XTE J1118+48. The top plot shows a simple model
consisting of a constant flux (at 100 counts/bin) multiplied by
the carbon absorption component of the optical blocking filter (OBF) model.Ê
The optical depth of the carbon component was increased by 13\% to
match the data in the 0.283-0.2865 keV region.Ê The data were then
corrected by this adjusted filter model to obtain the data in the
lower plot.Ê The residual is a sharp edge recovering smoothly
beyond 0.2870 keV.}
   \end{figure}

\begin{figure}
   \begin{center}
   \begin{tabular}{c}
   \includegraphics[height=8cm]{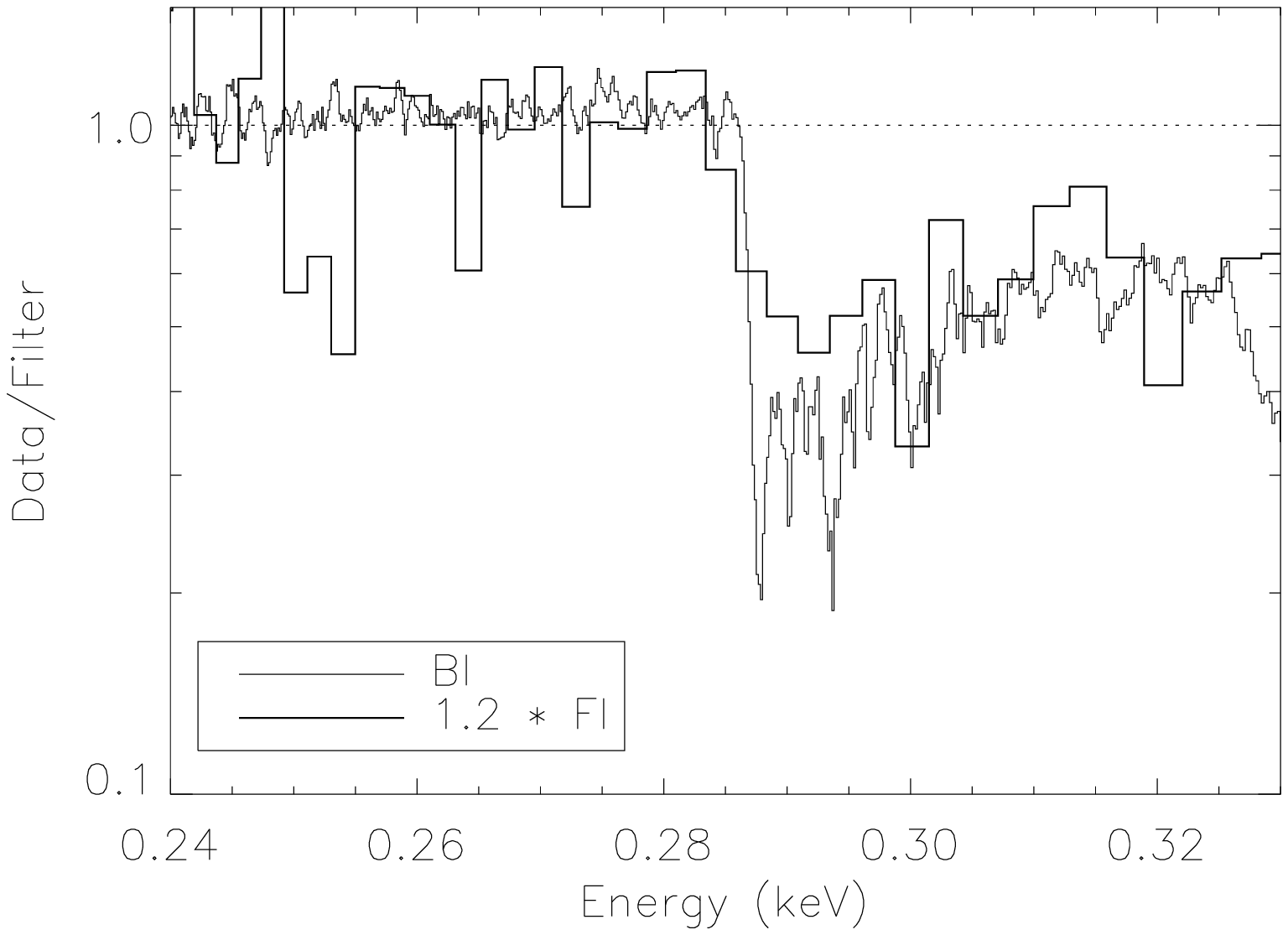}
   \end{tabular}
   \end{center}
   \caption
{ \label{fig:bifi} Data from the frontside-illuminatd (FI) chip on
the +1 LETGS order are compared to the spectrum from the -1
side, which is on a backside-illuminated (BI) chip.Ê The
coarsely binned data are from the FI chip in the C-K region where
the overall QE is quite low, so the data are heavily binned.
Nevertheless, there is a distinct edge that is not taken into
account in the OBF filter model.Ê The FI data agree
reasonably well with the BI results, after accounting for a
systematic QE difference of 20\%, except that the BI data are
somewhat lower than the FI data near 0.29 keV. }
   \end{figure}

In April 2000, data from a much brighter source, XTE J1118+480, were
obtained.Ê Because it was so bright at 0.28 keV, the
data were not rebinned, so that the filter model could be examined
in detail.  Fig.~\ref{fig:xtej1118} shows
the C-K region and how a slight adjustment to the ACIS filter
model in energy (0.5 eV) and in optical depth at C-K
(13\%) gives a good fit in the 0.283-0.287 keV
region (near edge structure) but that there is a strong sharp edge
above 0.287 keV.  No reasonable adjustment of the filter would be
able to explain the excess absorption above 0.287 keV.
Fig.~\ref{fig:bifi} shows that the edge is also present in the +1
order, where the C-K edge is observed on a frontside-illuminated
(FI) chip, which has significantly less quantum efficiency at 0.28 keV than
a backside-illuminated (BI) chip.  In these early observations, there
were no clear signs of other edges nor suspicion of
time variability, so a simple model consisting entirely
of a layer of carbon about 170 nm deep
was developed to correct for the C-K edge absorption.  An effective area
correction was released for {\em Chandra} LETGS analysis in 2001
(v2.6 of the {\em Chandra} calibration data base).


\section{Refining the Composition}

In early 2002, there was a clear indication of a progressive
drop in the ACIS low energy quantum efficiency\cite{plucinsky03}.
The C-K edge absorption that had been attributed only to LETG/ACIS
observations was now understood as a more general ACIS effect that
was changing with time.
A series of monitoring observations revealed more features as the
optical depth at C-K increased.  In June 2002, a 50 ks observation
of PKS 2155-304 revealed an O-K edge and possible residuals at the
F-K edge, as reported by
Plucinsky {\it et al.} (2003)\cite{plucinsky03}.  In October, 2002,
O-K and F-K edges were observed in a very long
observation of Mk 421 while it was in a very bright state.
A model of the absorption due to the contaminant was
developed as follows.

\subsection{Upgrading the C-K Edge Model}

\begin{figure}
   \begin{center}
   \begin{tabular}{c}
   \includegraphics[height=17cm]{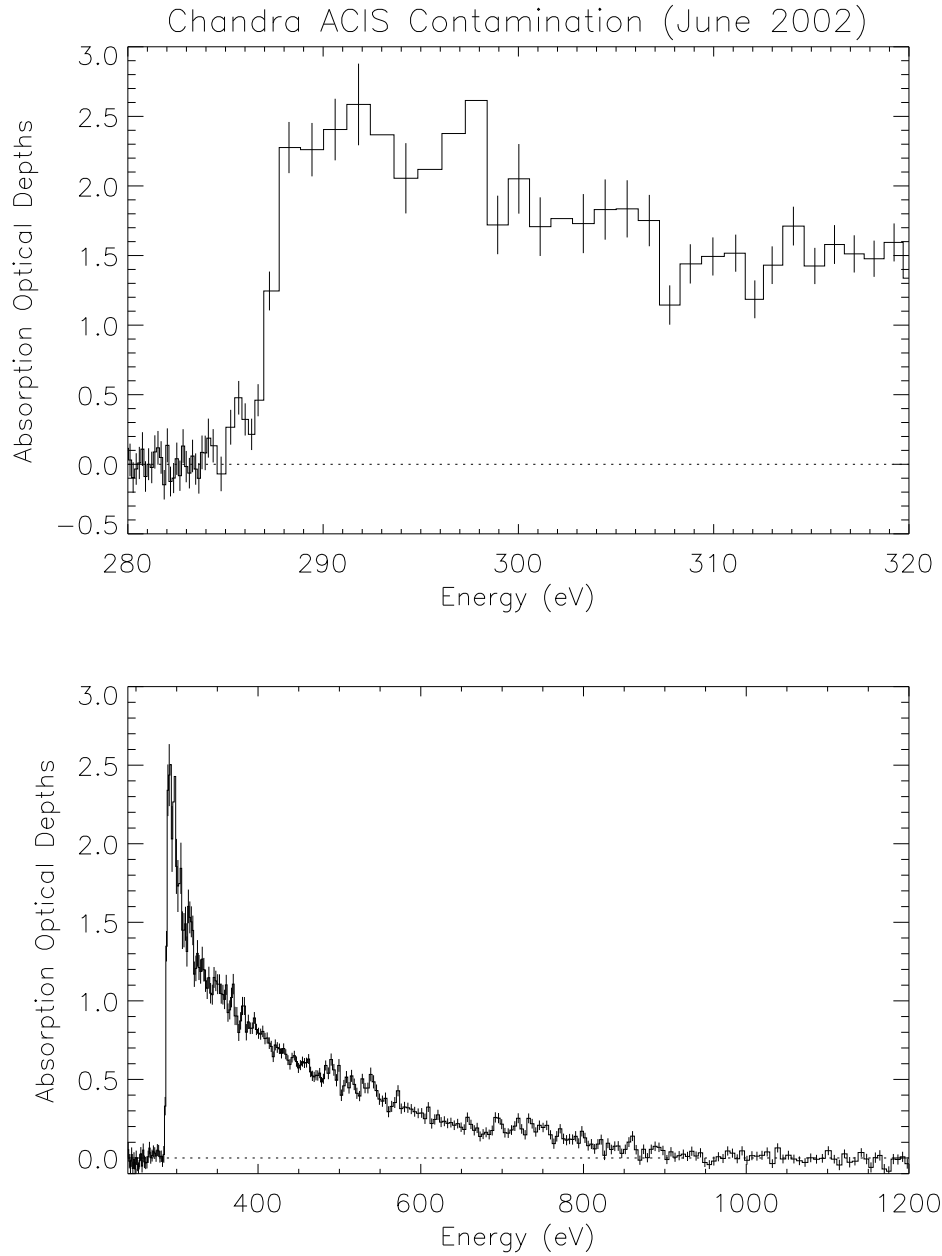}
   \end{tabular}
   \end{center}
\caption { \label{fig:contaminant} The optical depth of the
contaminant as a function of energy, derived by adaptively binning
the observed spectrum, $n_E$, fitting to a model, $N_E$, given by
eq.~\ref{eq:2pl} and
computing $\tau(E) = -\ln ( n_E/N_E )$.  There are several features
to notice.  There is a detectable opacity beteween 0.285 keV and
0.287 keV that was not obvious previously.  Above the edge at
0.287 keV, there is a monotonic decline except for some possible
structure near 0.5 keV (near the O-K edge) and 0.7 keV (near the
F-K edge). }
   \end{figure}

The observation of PKS 2155-304 in June, 2002 (observation ID 3669)
was analyzed in detail to obtain a good overall spectrum, $n_E$,
in units of ph cm$^{-2}$ s$^{-1}$ keV$^{-1}$.  We used a
continuum model of the form

\begin{equation}
\label{eq:2pl}
N_E = \frac{K_1 E^{-\Gamma_1}}{ 1 + \frac{K_1}{K_2} E^{\Gamma_2-\Gamma_1} }
\end{equation}

\noindent
where $N_E$ is the spectral flux in the same units
as $n_E$, $K_i$ are the normalizations
of the two components in the same units, and $\Gamma_i$ are
the spectral indices of the power law components.
In this model, there are no spectral breaks, only smooth
curvature at the point where the two power law components
have equal contributions.  Thus, no sharp features are added via
modeling.  Excluding the 0.28-1.0 keV region from the model fit
gives $N_E$ that is not affected by absorption.
The ratio of the data to the model expressed as an
optical depth due to absorption is computed as
$-\ln(n_E/N_E)$, plotted in Fig.~\ref{fig:contaminant}.

\begin{figure}
   \begin{center}
   \begin{tabular}{c}
   \includegraphics[height=8cm]{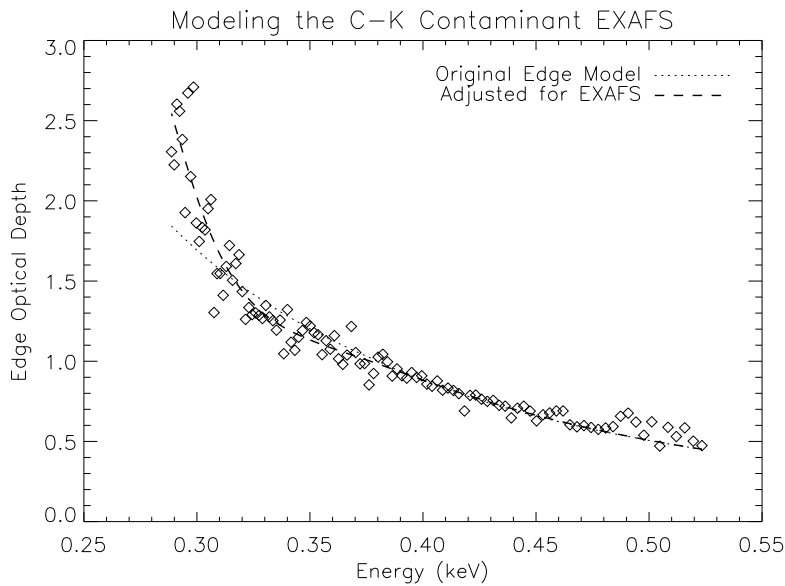}
   \end{tabular}
   \end{center}
\caption { \label{fig:ckexafs} Similar to the bottom half of
Fig.~\ref{fig:contaminant}
except the data have been rebinned to reduce statistical noise
and only the region being fit for extended fine structure
is shown.  The Henke model
of the C-K edge (short dashed line) is a good match to the data in
the 0.38 to 0.48 keV range, so the Henke optical constants are used
above 0.48 keV.  Below 0.38 and down to the edge, a damped
oscillation adjustment is used (eq.~\ref{eq:ripple}). }
   \end{figure}

\begin{figure}
   \begin{center}
   \begin{tabular}{c}
   \includegraphics[height=17cm]{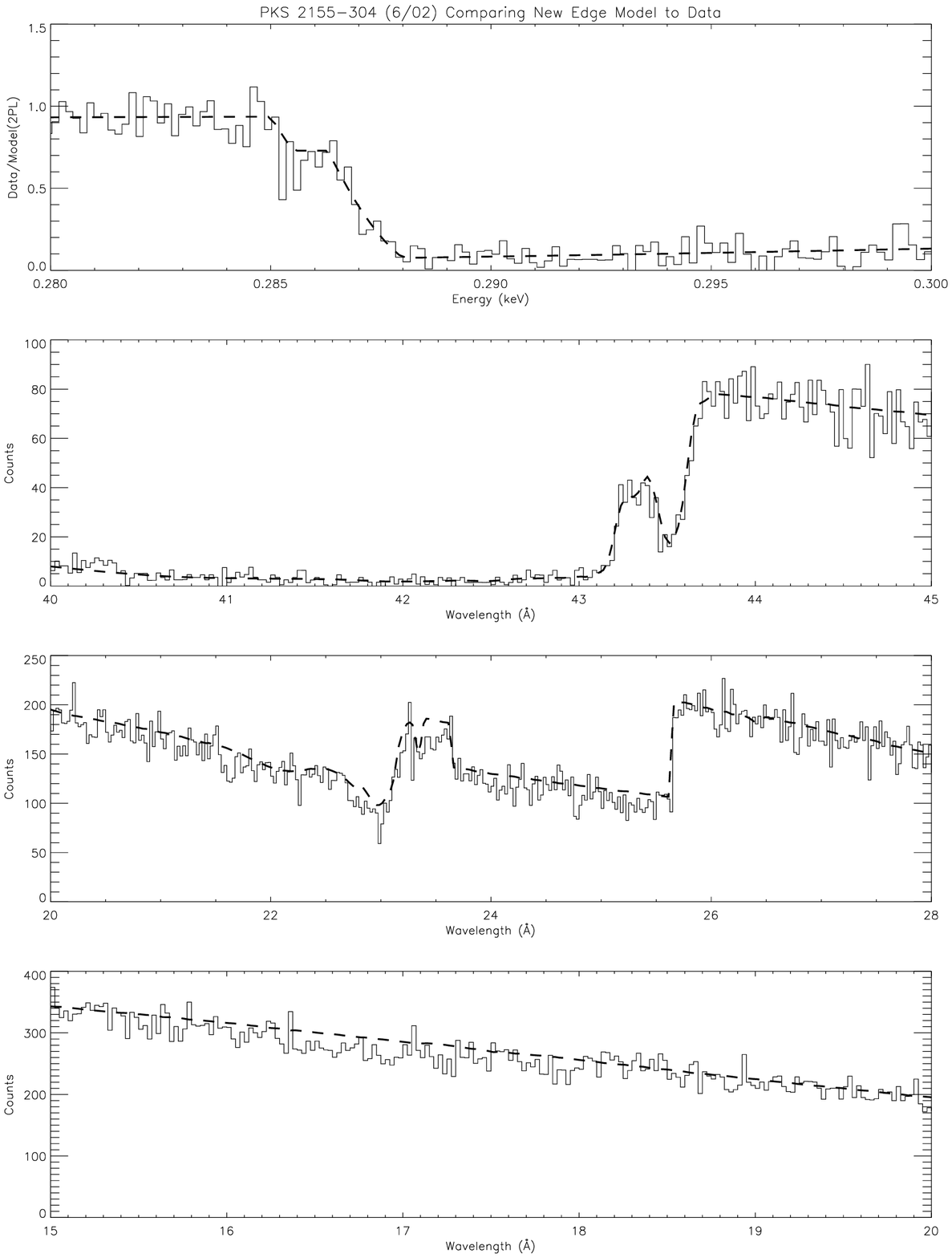}
   \end{tabular}
   \end{center}
\caption { \label{fig:ckexafs-data} Comparison of the new C-K edge model with
data from a long PKS 2155-304 observation.
The top panel shows the ratio of the data to the model without
including contamination so that the near edge structure is apparent and
the new model of the absorption due to
contamination is overlaid.  The remaining panels show the model with the
new absorption model folded with the effective area.  The data fit the
new model well with a slight systematic offset starting near 25.6 \AA\ where
there is a jump where one side goes from a BI chip to a FI chip, so the
systematic deviation from 20 to 26 \AA\ is primarily due to a slight error
in the correction to bring the QEs of these chips into agreement.  One
can also begin to see the 18 \AA\ (0.69 keV) edge due to
F-K that is modeled using a bright
Mk 421 observation and was ignored when fitting this C-K edge model.}
   \end{figure}

The data from Fig.~\ref{fig:contaminant} (bottom) were
further rebinned in order to model the
overall shape of the absorption (Fig.~\ref{fig:ckexafs}).
Fig.~\ref{fig:ckexafs} shows the new model for the C-K edge that starts with the
Henke constants from the Center for
X-ray Optics (CXRO\footnote{See
{\tt http://www.cxro.lbl.gov/}.}) except that the optical depth was
multiplied by a damped ripple adjustment factor:

\begin{equation}
\label{eq:ripple}
1 + A \exp( -x/x_d ) \cos( f x )
\end{equation}

\noindent
where $x \equiv (\lambda_{C-K} - \lambda) / \lambda_{C-K}$ and
$\lambda_{C-K}$ is the wavelength of the edge: 43.25 \AA\ (0.2867 keV).
The constants of the
adjustment were estimated (chi-by-eye): $A = 0.4$, $x_d =
4.0/\lambda_{C-K}$, and $f = 2 \pi \lambda_{C-K}/16$.  These provided a sharp
increase in the opacity just to the high energy side of the edge which damps out
to higher energies
so that it is asympotically equivalent to the form from the CXRO
web site.
The oscillatory term accounts for the ``ripple'' of the deviation.
The near edge X-ray absorption feature near 0.286 keV was added to the model
empirically as a new
shallow edge.  This shallow edge was actually detected in the observation
of XTE J1118+48 (Fig.~\ref{fig:xtej1118}) but was modeled by a 13\%
adjustment of the filter; upon increasing to an optical depth of 30\%,
it was clear that this feature was part of the C-K edge structure of
the contaminant.  Thus, the 13\% filter
adjustment was not applied when developing the C-K edge model upgrade.
The new model of the C-K edge is folded with the
instrument response and shown against the PKS 2155-304 data in
Fig.~\ref{fig:ckexafs-data} and clearly shows the small extra
absorption in the 0.285-0.287 keV region.

\subsection{Modeling the F-K and O-K Edges}

\begin{figure}
   \begin{center}
   \begin{tabular}{cc}
   \includegraphics[height=6cm]{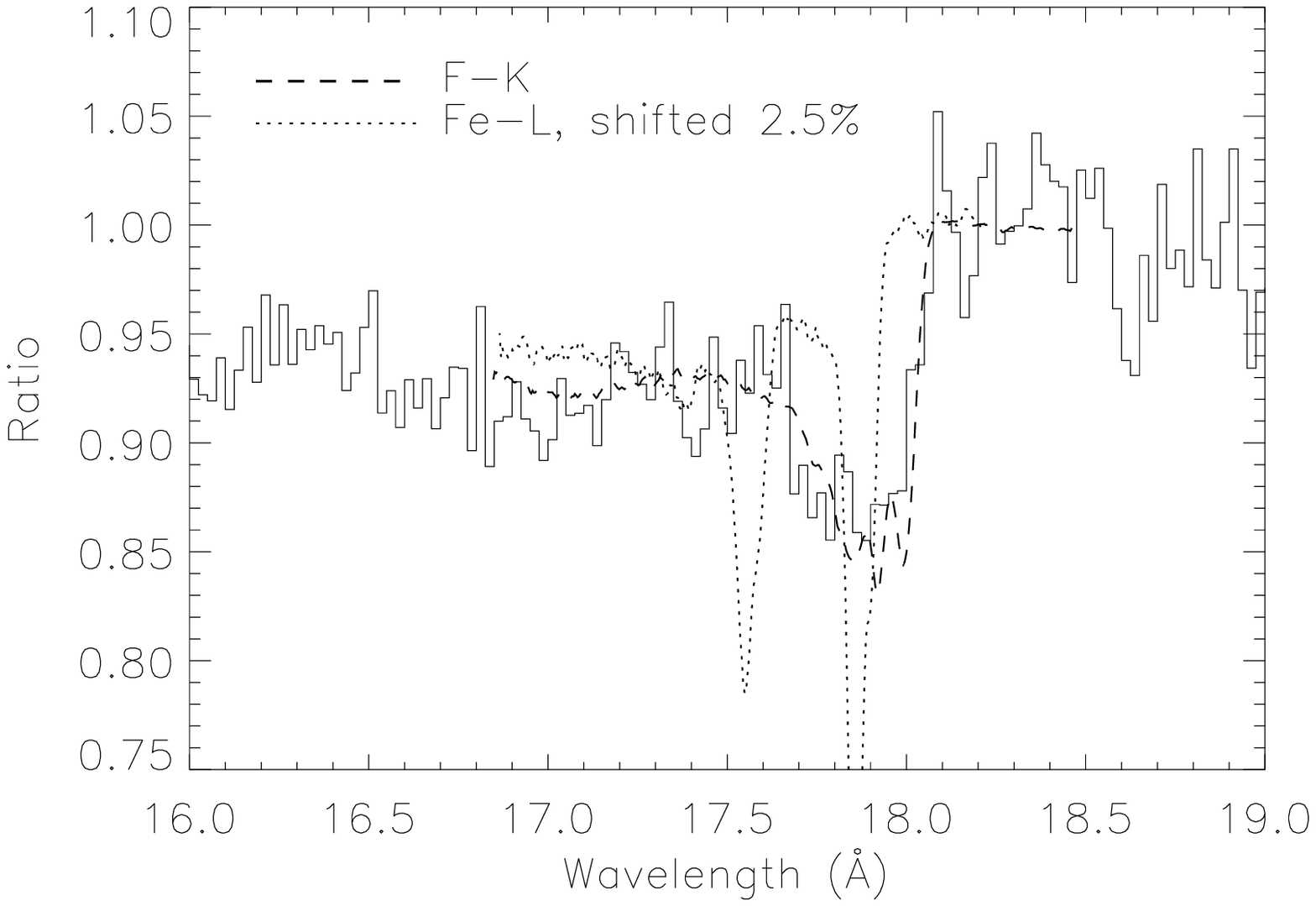}
   \includegraphics[height=6cm]{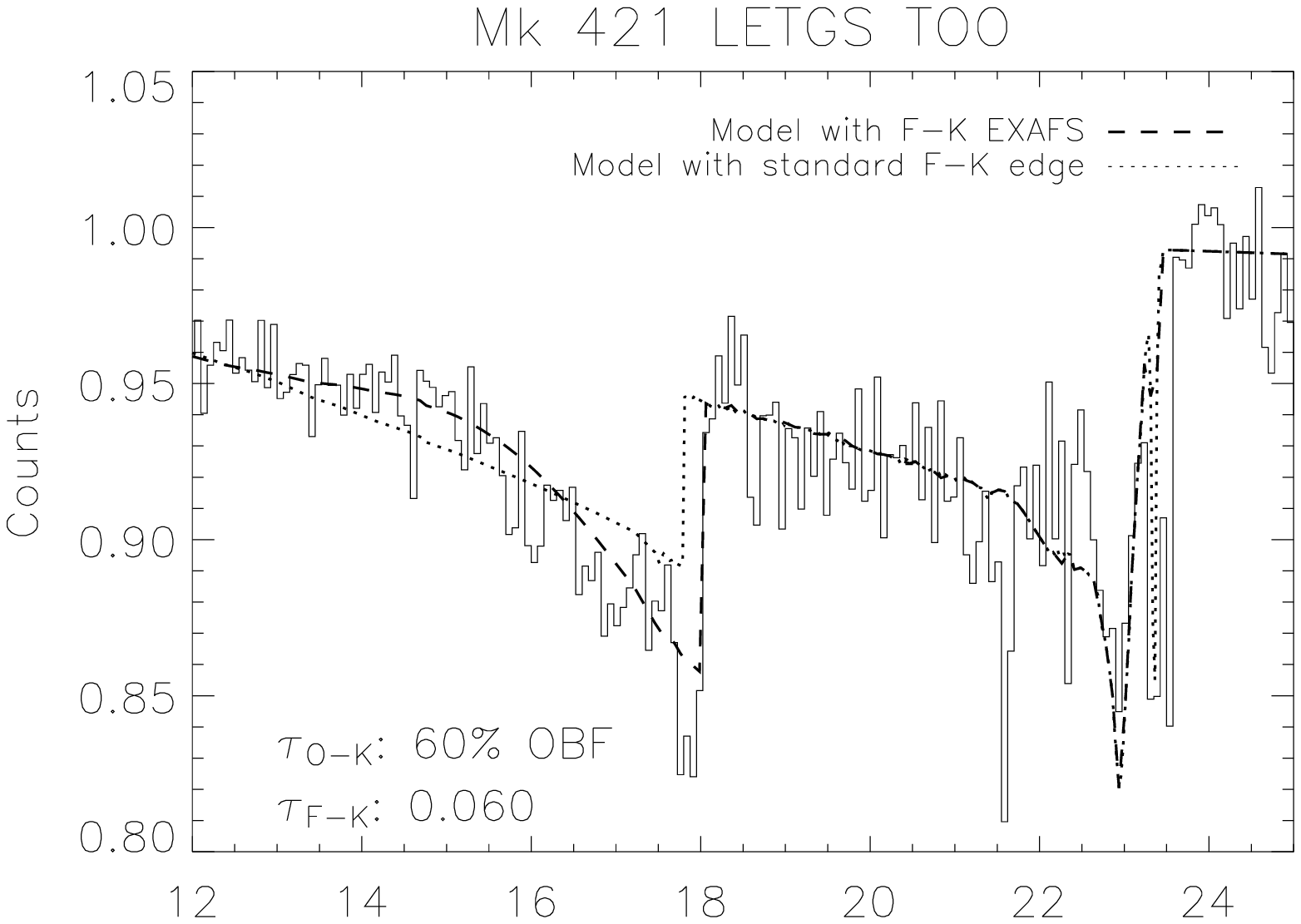}
   \end{tabular}
   \end{center}
   \caption 
   { \label{fig:fkmodel} 
{\bf a)} (Left) The 18 \AA\ (0.688 keV) edge is compared to two models.  The
dotted line represents possible Fe-L absorption, redshifted to
$z = 0.025$.  The lines are too narrow to fit the data.  The F-K edge,
however, fits very well without shifting.  F-K and Fe-L opacity
data were obtained by measuring the X-ray transmission of perfluoro-adamantine
and butadiene iron tricarbonyl with a scanning transmission X-ray
microscope\cite{stxm2003}.
{\bf b)} (Right) Modeling the 12-25 \AA\ (0.5-1 keV) region
of the Mk 421 X-ray spectrum.
The O-K absorption is dominated by a
resonance feature at 23.35 \AA\ (0.531 keV) and a deep
edge at 23.0 \AA\ (0.540 keV).  Other
narrow features in this region are associated with the source or
the interstellar medium.  The 18 \AA\ (0.69 keV) edge is modeled using Henke
optical constants for fluorine with a modification factor given
by eq.~\ref{eq:ripple} and shifting the edge to 18.03 \AA\ (0.688 keV).
}
   \end{figure}

As with PKS 2155-304, we extracted the
observed count spectrum, $n_E$, for the Mk 421 observation in October
2002 (obs ID 4148, provided by F. Nicastro).
This time, the C-K edge model was included with the
model of $N_E$, allowing the depth of the edge to vary using a
single depth parameter.  The optical depth due to absorption is computed
as before and is plotted in Fig.~\ref{fig:fkmodel}. The structure of the
O-K edge is very similar to the O-K edge found in the ACIS optical
blocking filter.  The model was taken from a spectral decomposition of
the OBF\cite{chartas1996} and then matched
to the Henke optical constants in the 18-21
\AA\ region for extrapolation shortward of 18 \AA.
This model has a resonance
feature that shows up as a deep narrow absorption line near 23.3 \AA\ that
is clearly detected.

\begin{figure}
   \begin{center}
   \begin{tabular}{c}
   \includegraphics[height=11cm]{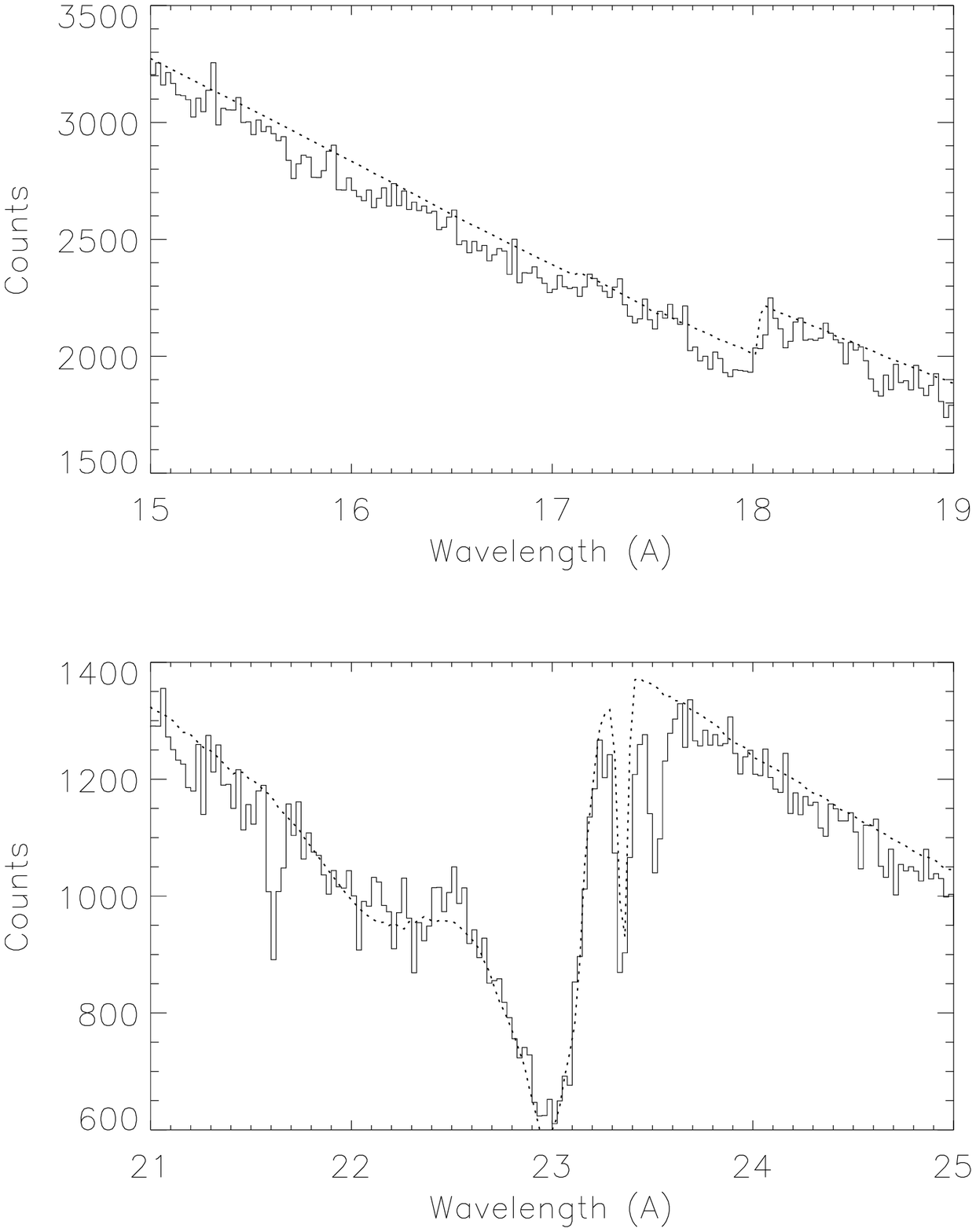}
   \end{tabular}
   \end{center}
   \caption 
   { \label{fig:fkdata} 
Count spectra at the edges of F-K (top) and O-K
(bottom).  The overall structure is modeled to better than 3\% and the
only remaining
features are intrinsic to the source (or to the gas between the target
and the telescope, see Nicastro et al. 2003, in
preparation).  There is a slight normalization error in the
F-K edge region due to the systematic effects of correcting for the
relative error in the BI QEs compared to the FI QEs\cite{marshall03}.}
   \end{figure}

The edge at 18 \AA\ (0.69 keV) is attributed to the fluorine K edge.
Fig.~\ref{fig:fkmodel}a shows that the edge
cannot be identified with redshifted Fe-L.
The F-K edge extended fine structure was
then fit by a similar method used to fit the C-K.
First, the Henke optical constants were obtained from the
CXRO.  Then, the edge was positioned at $\lambda_{F-K}$ = 18.03
\AA\ (0.688 keV) and fit using eq.~\ref{eq:ripple} but substituting
$\lambda_{F-K}$ for $\lambda_{C-K}$.
The constants of the adjustment
were estimated: $A = 0.6$, $x_d = 4./\lambda_{F-K}$, and $f =
2 \pi \lambda_{F-K}/8.5$.  Note the similarity of the parameters to that used in
fitting the C-K edge.   The result is shown in Fig.~\ref{fig:fkmodel}b.
The overall fit to the Mk
421 data in the 0.22-1.0 keV region is shown in Fig.~\ref{fig:fkdata}.

\subsection{Elemental Abundances}

The resulting spectral model of the contaminant
absorption is available in a file at the
HETGS calibration web site.\footnote{See
{\tt http://space.mit.edu/CXC/calib/contamination\_model.txt}.}  There
are four columns, corresponding to wavelength (\AA) and the remaining
three correspond to the absorption in optical depths due to carbon, oxygen, and
fluorine.  The fits gave (Henke equivalent) edge optical depths of 2.09
$\pm$ 0.02, 0.07 $\pm$ 0.03, 0.100 $\pm$ 0.007, and 0.066 $\pm$ 0.005 for the
C-K, N-K, O-K and F-K edges, respectively, for the reference date
of JD 2452574.0.
We obtain the following column densities for C, O, and F:
2.0$\times 10^{18}$, 1.75$\times 10^{17}$, and
1.45$\times 10^{17}$ atoms cm$^{-2}$.  For N-K, the PKS 2155-304 data
provided a better upper
limit of the edge optical depth of about 5\%, giving a column density of
7.1$\times 10^{16}$ atoms cm$^{-2}$.  Carbon is more abundant than the other
elements by factors of 11.5, 14, and $>$30 for F, O, and N, respectively. 
The uncertainties on these ratios are about $\pm$ 1.
Using this model, the mass surface density as of early August 2003 is
56 $\mu$g cm$^{-3}$.  For an estimated density of 1.5 g cm$^{-3}$, the
contaminant should be about 370 nm thick.  For comparison, the thickness
of the ACIS OBF is 330 nm.

\section{Comparing to Transmissions of Known Compounds}

Detecting F-K in the absorption spectrum indicates that some portion
of the contaminant is related to on-board fluorocarbon-based
compounds such as Braycote 601 and Krytox, which are used in the
{\em Chandra} mechanisms.  However,
the abundance ratios measured in the contaminant
are not found in any of the materials actually
aboard the spacecraft.  The contaminant is much more carbon rich than
any known fluorocarbon.  Therefore, we are not detecting these
compounds directly.

\begin{figure}
   \begin{center}
   \begin{tabular}{c}
   \includegraphics[height=10cm]{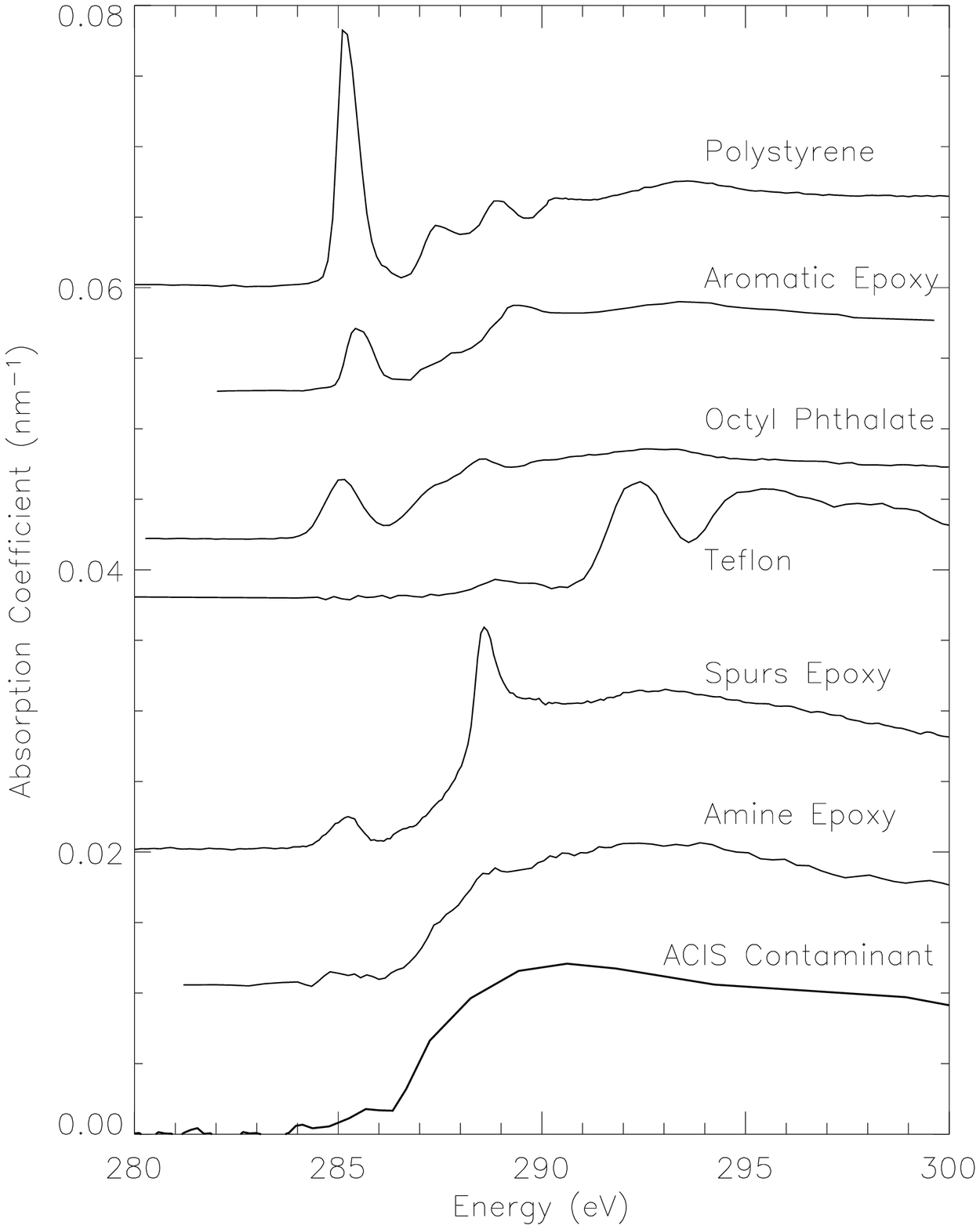}
   \end{tabular}
   \end{center}
   \caption { \label{fig:compare-compound}
Comparison of the opacity
of the ACIS contaminant near the C-K edge to those for 6
different materials measured using a scanning transmission X-ray
microscope\cite{stxm2003}.  The intensity scales of
the reference opacities are
quantitative linear absorption per nm of material; that for the
contaminant has been scaled to have a similar amplitude. Arbitrary
offsets are used for clarity. The absorption spectra of compounds
containing C=C or C=O double bonds or C-F single bonds do not
match the spectra of the contaminant.  A good match is found to an
aliphatic amine epoxy.}
   \end{figure}

Another clue to the origin of the contaminant comes from the structure
of the absorption near the C-K edge.  In
fig.~\ref{fig:compare-compound}, we compare the absorption spectrum
of the ACIS contaminant near the C-K edge to
those of a number of hydrodcarbon and fluorocarbon species similar
to species used in Chandra.
The sharp peaks at 0.285 keV are
associated with C=C double bonds present in unsaturated and
aromatic molecules.
Fluorocarbons like teflon, Braycote and Krytox
have sharp structures in the 0.289-0.296 keV range associated with
C-F bonds.
The ACIS contaminant shows neither of these features.
Aliphatic hydrocarbons (those containing only of mainly C-C single
bonds) have a C-K edge which is broad with relatively little
structure, similar to that observed for the ACIS contaminant.
We conclude that the contaminant is predominantly aliphatic;
i.e., consisting mostly of hydrocarbons.

We surmise that the contamination material is actually a product
of radiation damage of fluorocarbon based materials.
C-F bonds are
readily broken by X-rays or cosmic rays leading to loss of
fluorine, which flies off as FH,ÊF$_2$, or a small
fluorocarbon.
If the radiation-damaged fluorocarbon remains on
the detector or filter surface, the damage event leaves behind a carbon-rich
residue, which could bind quite strongly and build up over time.

\section{Time Dependence}

\begin{figure}
   \begin{center}
   \begin{tabular}{cc}
   \includegraphics[height=6cm]{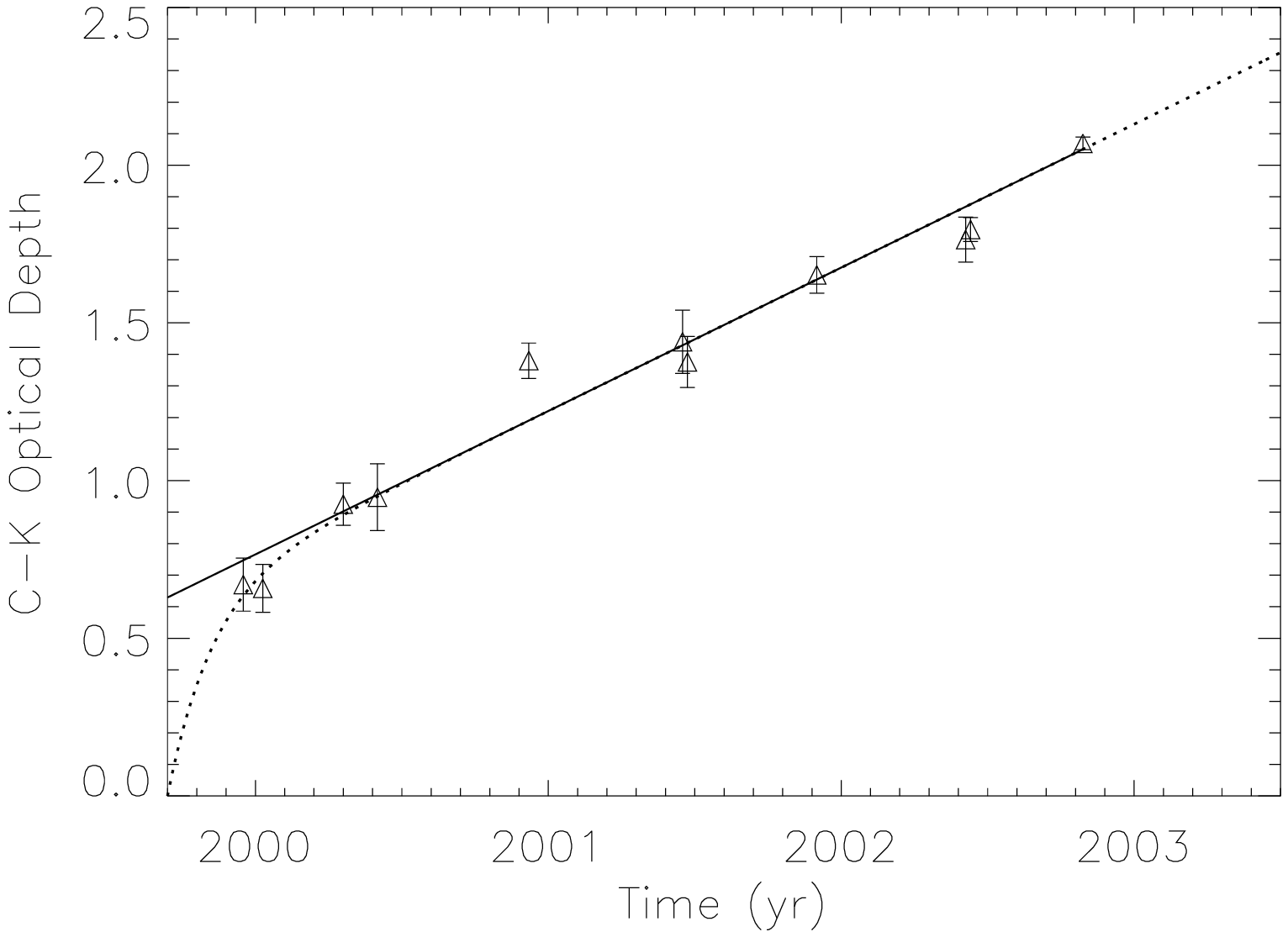}
   \includegraphics[height=6cm]{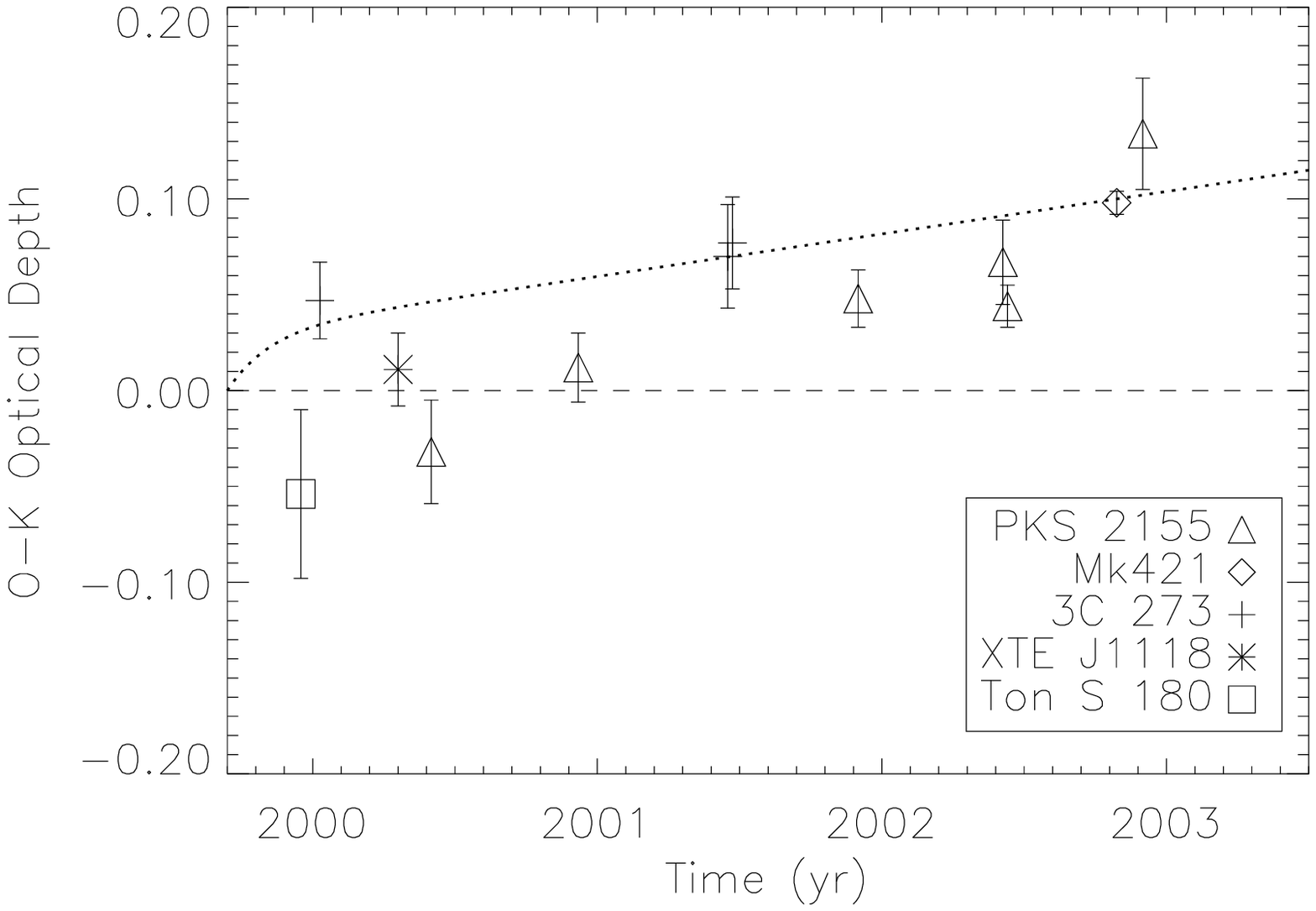}
   \end{tabular}
   \end{center}
   \caption 
   { \label{fig:edge-vs-time} 
{\bf a)} (Left) The C-K edge optical depth as a function of time for 11
observations of essentially featureless sources observed with the LETG
and ACIS.  The solid line is a linear regression that is not forced to
go through zero, while the dashed line is a model with similar
asymptotic behavior that is forced to zero at ACIS door opening.
{\bf b)} (Right) O-K edge measurements as a function of year compared to
the model for the time dependence of the O-K optical depth.  The model
assumes that the O-K optical depth scales with the C-K edge and is
forced to go through the best data point (Mk 421, in late 2002).  In
general, the trend given by the model is matched by the data: the O-K
edge is detected only in the last year or so and is smaller in the first
year.  In particular, the possibility that the O-K edge is nearly
constant after 2000.0 seems to be ruled out. }
   \end{figure}

Using the contamination model derived above, the LETG/ACIS data were fit
for many sources with smooth spectra, allowing the C-K, O-K, and F-K
edges to vary independently.  The results are
shown in Fig.~\ref{fig:edge-vs-time}.
Then the C-K edge depth was fit to a simple time
dependence model. The data are well fit by a linear model but a perhaps
more physical model was adopted to allow the optical depth to be zero at
the time of the opening of the ACIS door.  The equation for the time
dependence is

\begin{equation}
\tau = a(t-y_0) - a(t_0-y_0) e^{\frac{t-t_0}{q}},
\end{equation}
\noindent
where $\tau$ is the C-K optical depth, $a$ is the asymptotic rate of increase
(0.455 opt. depths per year), $t_0$ is the date (1999.70) at which $\tau$ is
forced to be zero, $y_0$ is the year (1998.32) at which the linear
regression gives $\tau = 0$, and $q$ is the estimated timescale on which the
optical depth approaches the asymptotic form: 0.15 yr.

The shape of the time dependence was assumed to be
the same for each of the three detected edges and were merely scaled to
observed optical depths (see the previous section) for the date of the
Mk 421 observation (year 2002.86).  Because the optical depth values used to
create the contamination model were slightly different -- 2.10, 0.09, and
0.06 (for C-K, O-K, and F-K, respectively) -- the time factors for each
component were merely rescaled.  The resulting time dependence file is
available on-line.\footnote{See
{\tt http://space.mit.edu/CXC/calib/time\_factor.txt}.}

The O-K edge data do not fit the model as well as expected for reasons
that are not completely understood.
Fig.~\ref{fig:edge-vs-time} shows that the O-K edge
measurements are mostly below the model.
These deviations
are not readily explained by O-K in the interstellar medium (ISM).
In all the PKS 2155
measurements, for example, the $N_H$ was free and varied between 1.0 and
1.7$\times 10^{20}$ cm$^{-2}$.  The ``nominal'' value is about
1.35$\times 10^{20}$ cm$^{-2}$, for
which two different ISM models give an O-K edge of 0.054 optical depths.
If the $N_H$ were systematically decreased to
1.0$\times 10^{20}$ cm$^{-2}$, say, then the
average O-K due to the contaminant would increase by 0.027 -- which
is insufficient to explain the difference between the model and the
data.

\section{Comparison to the External Calibration Source}

\begin{figure}
   \begin{center}
   \begin{tabular}{c}
   \includegraphics[height=8cm]{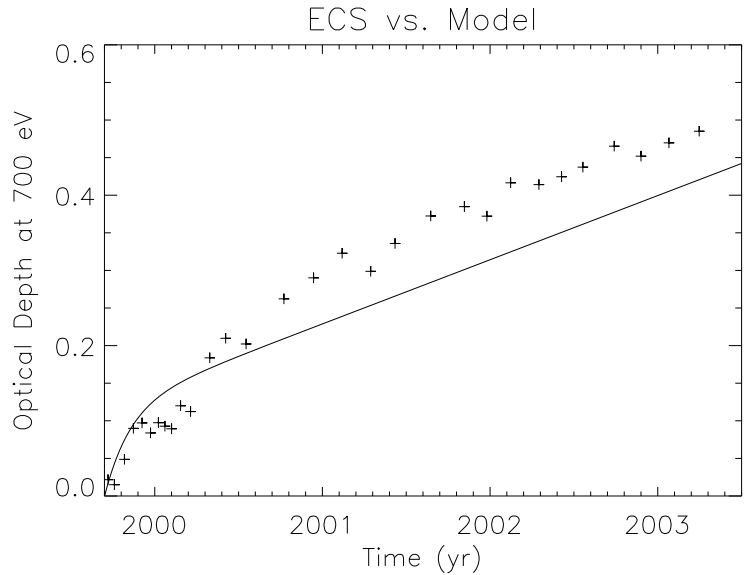}
   \end{tabular}
   \end{center}
\caption { \label{fig:ecs} The optical depth at 0.70 keV
determined from the ratio of the fluxes in the external cal
source (ECS)
Mn L and Fe L lines to the flux of the Mn K line integrated
over the entire S3 (BI) chip as a function
of time for the first 3.5 years of {\em Chandra} operations.
Also shown is the
model of the time dependence derived from fits to the C-K
edges (see fig.~\ref{fig:edge-vs-time}).  The contamination
model is 20\% lower than the computed optical depth over the
last two years.  The difference is about 0.1 optical depths,
corresponding to a 10\% difference in total transmission.
}
   \end{figure}

Plucinsky {\it et al.}\cite{plucinsky03} showed how the ratio of the
L line fluxes in the External Calibration Source (ECS) relative to
the Mn K line flux is a monotonically decreasing function of time.
This flux ratio appears to be a direct measure of the transmission
of the contaminant at the energies of the lines in the L line
complex.  Several lines are expected, dominated by the $\alpha_1$
and $\alpha_2$ lines of Fe and Mn at 705.0 and 637.4 eV, respectively.
The F-K edge falls between these energies and it is a fortuitous
circumstance that the transmission through the contamination
absorption model developed here is very nearly the same at these
two energies, so we choose $E = 700$ eV at which to evaluate the
absorption model for comparison to the ECS data, shown in
Fig.~\ref{fig:ecs}.  The L/K ratios have been converted to optical
depths for ease of comparison with the contamination model.

Fig.~\ref{fig:ecs} shows that the contamination model
developed here explains
about 80\% of the opacity that absorbs the ECS L lines.
We cannot as yet explain what causes the remaining 20\%, which
is 0.1 optical depths at the present time
(mid-2003).  Thus, using this
contamination model may result in systematically low fluxes
by 10\% in observations if the ECS gives an accurate portrayal
of the absorption on the detector focal plane.  Due to the
relatively high
temperature expected for the ECS, it seems unlikely that
sufficient contaminant has condensed on the ECS
itself that might explain this difference.
One possibility is that the radiation-generated molecular
fragments have a range of sizes that would result in a range
of condensation temperatures, so that the ECS-LETGS differences
might provide an estimate of the fraction of fragments
with high condensation temperatures.

One class of attempts to explain the extra 0.1 optical depth at 700 eV
involve adding some other material in the contaminant
but most of these would have
clear spectral edges at higher energies (in the case of Si and Mg).
Hydrogen wouldn't show an edge in the LETGS band but the
column density that would be required to give an optical depth of
0.1 at 700 eV is 2.7 $\times 10^{21}$ cm$^{-2}$, or 1300 atoms
of H per atom of C.  It would seem to be an unlikely mixture
that would somehow adhere to the ACIS detector.  One interesting
possibility is Au, used as a coating in much of ACIS.
It would require only a 50 nm layer to
provide an opacity of 10\% at 700 eV and yet the M edge at 2.3 keV
would have a depth of only 2\%.  Such an edge is difficult to measure
at present, given the strong edge due to Au in the gratings
themselves.

\section{Further Work}

We have obtained X-ray transmission data for a sample of Braycote
and found that the material was quite sensitive to irradiation.
We are preparing results on the x-ray transmission of the
radiation-damaged oil.  Preliminary results indicate that the
X-ray transmission changes character in the manner expected:
the 0.289 keV opacity maximum shifts downward to 0.287 keV, where
it is found in the {\em Chandra} system.  We will also test for
changes at the F-K edge.

We have recent evidence that the contaminant opacity varies somewhat
from center to edge.  Although spatial nonuniformity affects the
comparison of the ECS and LETGS results, the discrepancy between
these estimates of the opacity at 0.7 keV 
(fig.~\ref{fig:ecs}) remains unexplained.

\acknowledgments     
 
This work was supported in part by contracts SAO SV1-61010 and
NAS8-39703.  We thank Dan Dewey for useful comments on the
manuscript.


\bibliography{contaminant}   
\bibliographystyle{spiebib}   

\end{document}